\documentclass[twocolumn,times]{aastex631}

\usepackage{natbib}
\usepackage{epsfig,amsmath,amssymb}
\usepackage{graphicx}
\usepackage{xcolor}
\usepackage{bm}

\def\CII{[\ion{C}{2}]}
\def\kms{km\,s$^{-1}$}
\def\lya{Ly-$\alpha$}

\begin{document}

\title{\Large  ALMA 400 pc Imaging of a $\bm{z=6.5}$ Massive Warped Disk Galaxy}

\correspondingauthor{Marcel Neeleman}
\email{mneeleman@nrao.edu}

\author[0000-0002-9838-8191]{Marcel Neeleman}
\affiliation{National Radio Astronomy Observatory, 520 Edgemont Road, Charlottesville, VA, 22903, USA}
\affiliation{Max-Planck-Institut f\"{u}r Astronomie, K\"{o}nigstuhl 17, D-69117, Heidelberg, Germany}
\author[0000-0003-4793-7880]{Fabian Walter}
\affiliation{Max-Planck-Institut f\"{u}r Astronomie, K\"{o}nigstuhl 17, D-69117, Heidelberg, Germany}
\author[0000-0002-2662-8803]{Roberto Decarli}
\affiliation{INAF -- Osservatorio di Astrofisica e Scienza dello Spazio di Bologna, via Gobetti 93/3, I-40129, Bologna, Italy}
\author[0000-0002-0174-3362]{Alyssa B. Drake}
\affiliation{Centre for Astrophysics Research, Department of Physics, Astronomy and Mathematics, University of Hertfordshire, Hatfield AL10 9AB, UK}
\author[0000-0003-2895-6218]{Anna-Christina Eilers}
\affiliation{MIT Kavli Institute for Astrophysics and Space Research, 77 Massachusetts Avenue, Cambridge, MA, 02139, USA}
\author[0000-0001-5492-4522]{Romain A. Meyer}
\affiliation{Max-Planck-Institut f\"{u}r Astronomie, K\"{o}nigstuhl 17, D-69117, Heidelberg, Germany}
\affiliation{Department of Astronomy, University of Geneva, Chemin Pegasi 51, 1290 Versoix, Switzerland}
\author[0000-0001-9024-8322]{Bram P. Venemans}
\affiliation{Leiden Observatory, Leiden University, PO Box 9513, 2300 RA Leiden, The Netherlands}

\begin{abstract}
We present $0\farcs075$ ($\approx$400 pc) resolution ALMA observations of the \CII\ and dust continuum emission from the host galaxy of the $z = 6.5406$ quasar, P036$+$03. We find that the emission arises from a thin, rotating disk with an effective radius of $0\farcs21$ (1.1 kpc). The velocity dispersion of the disk is consistent with a constant value of $66.4 \pm 1.0$ \kms, yielding a scale height of $80 \pm 30$ pc. The \CII\ velocity field reveals a distortion that we attribute to a warp in the disk. Modeling this warped disk yields an inclination estimate of $40.4 \pm 1.3 \degr$ and a rotational velocity of $116 \pm 3$ \kms. The resulting dynamical mass estimate of $(1.96 \pm 0.10) \times 10^{10} M_\odot$ is lower than previous estimates, which strengthens the conclusion that the host galaxy is less massive than expected based on local scaling relations between the black hole mass and the host galaxy mass. Using archival MUSE \lya\ observations, we argue that counterrotating halo gas could provide the torque needed to warp the disk. We further detect a region with excess (15-$\sigma$) dust continuum emission, which is located 1.3 kpc northwest of the galaxy's center and is gravitationally unstable (Toomre-Q $<$ 0.04). We posit this is a star-forming region whose formation was triggered by the warp, because the region is located within a part of the warped disk where gas can efficiently lose angular momentum. The combined ALMA and MUSE imaging provides a unique view of how gas interactions within the disk-halo interface can influence the growth of massive galaxies within the first billion years of the universe.
\end{abstract}


\section{Introduction}
\label{sec:intro}
In our current hierarchical galaxy assembly paradigm, galaxies are thought to grow through mergers and through accretion of gas from the cosmic web. The latter is believed to be especially important at high redshift, because it provides a way to funnel cold gas into the center of the galaxy, which can provide direct fuel for star-formation \citep[e.g.,][]{Keres2005,Dekel2009,Walter2020}. Detailed numerical simulations indicate that when these streams of gas get deposited onto the galaxy, they exchange angular momentum that disturb the interstellar medium (ISM) of the galaxy and thereby trigger the formation of stars \citep{Danovich2015}. Understanding the process of gas accretion from the halo onto the ISM of the galaxy (in the so-called disk-halo interface) is therefore crucial in order to understand how galaxies acquire their mass, form stars and evolve through cosmic time. 

Unfortunately the disk-halo interface is both difficult to model with numerical simulations as well as observe at the redshifts when most of the accretion is thought to occur. Simulations are hampered by the large range in spatial scales that needs to be captured in order to accurately model the physical processes that influence the accretion process \citep[e.g.,][]{Hummels2019}. Observations, on the other hand, are unable to detect the tenuous gas directly. In the few cases that a background quasar intersects the halo close to a galaxy, the gas can be studied in absorption, but these observations mostly probe just a single line-of-sight in the halo and therefore lack spatial information \citep[e.g.,][]{Tumlinson2017}.

Observations of the \lya\ emission line could provide a solution to observing the disk-halo interface at high redshift. Previous observations have shown that \lya\ emission extends well beyond the ISM of galaxies \citep[e.g.][]{Steidel2011}, and using the Multi-Unit Spectroscopic Explorer (MUSE) on the ESO-VLT, multiple studies have shown the feasibility of detected \lya\ emitting halos around individual high redshift galaxies \citep[e.g.,][]{Wisotzki2016,Leclercq2017}. By exploring the kinematics of the \lya\ emission line, these studies further show that the \lya\ halo is kinematically distinct from the galaxy's ISM \citep{Leclercq2020}, although the nature of \lya\ emission, which can both be scattered as well as produced through fluorescence \citep[e.g.,][]{Dijkstra2014,Mas-Ribas2016}, hampers the interpretation of the results. 

Extending these MUSE observations to high redshift quasars has revealed that \lya\ halos are ubiquitous around quasars at $z \sim 3$ \citep{Borisova2016,Arrigoni-Battaia2019}. This prompted the search for \lya\ halos around the highest redshift quasars \citep[$z > 6$;][]{Drake2019,Farina2019}. These studies have revealed that roughly a third of all $z > 6$ quasars have a bright ($L_{\rm Ly\alpha} \gtrsim 10^{43}$ erg s$^{-1}$) \lya\ halo and that these halos are also kinematically decoupled from the kinematics of the host galaxy \citep{Drake2022}. Recent simulations propose that these bright \lya\ halos are dense gas being deposited onto the galaxy that is illuminated by \lya\ from the galaxy's nucleus \citep{Costa2022}. This provides a theoretical backing for using \lya\ emission as a tracer of the gas kinematics in the disk-halo interface of high redshift quasars (with the caveats mentioned above). 

Studying the gas exchange between the halo and the disk in the host galaxies of $z > 6$ quasars is particularly interesting for several reasons. First, luminous quasars are believed to trace the most massive galaxies in the universe \citep[e.g.,][]{Volonteri2006,Overzier2009}. These galaxies are believed to form rapidly within dense environments, and observing them at $z > 6$ allows us to explore their gaseous environs when the galaxies were actively accreting material. From dynamical mass estimates and local scaling relations between the black hole mass and the host galaxy mass \citep{Ferrarese2000,Gebhardt2000,Kormendy2013,Reines2015}, we know that $z > 6$ luminous quasars have host galaxies that are under-massive compared to the mass of the black hole \citep[e.g.,][]{Neeleman2021}. This and their extreme star formation rates \citep[e.g.][]{Venemans2020} at least qualitatively suggests that the galaxies are still acquiring the bulk of their mass, either through mergers or gas accretion from the cosmic web. Second, we know that the host galaxies of $z > 6$ quasars have large quantities of centrally located cold gas  \citep[e.g.,][]{Walter2022}, suggesting that angular momentum loss plays a crucial role in the formation of these galaxies. Observing the gas surrounding galaxies can help us understand the cause of this angular momentum loss. Finally, for $z > 6$ quasar host galaxies we can study the interstellar medium in exquisite detail (resolutions of only a few 100 parsec) in the far-infrared using the Atacama Large Millimeter/sub-millimeter Array (ALMA); in particular using the singly ionized carbon far-infrared emission line \citep[\CII;][]{Venemans2019,Walter2022}.

In this paper we provide an in--depth look at one such luminous $z>6$ quasar, P036$+$03, which has J2000 coordinates of 02:26:01.88, $+$03:02:59.2. This quasar was part of the sample of eight luminous $z>6$ quasars discussed in \citet{Drake2022}, and has archival MUSE \lya\ observations \citep{Farina2019}. This paper describes new ALMA observations of the \CII\ and far-infrared dust continuum emission at a resolution of $0\farcs075$ ($\approx$ 400 pc). Previous $\approx 0\farcs14$ (0.75 kpc) resolution imaging indicated that the host galaxy of P036$+$03 is a disk \citep{Neeleman2021}. With the higher resolution imaging we can explore, in detail, the kinematics and distribution of gas and dust in the ISM of the galaxy, and by comparing these observations with the archival \lya\ emission, we can study the interaction between the halo gas and the disk.

This paper is organized as follows. In Section \ref{sec:obs} we describe the new ALMA observations. In Section \ref{sec:analysis} we present the analysis of the ALMA data, focussing on the kinematic modeling of the \CII\ emission and the comparison between the dust and continuum emission. In Section \ref{sec:discussion} we discuss the analysis, which we summarize in Section \ref{sec:conclusions}. Throughout this paper we use a standard flat $\Lambda$ cold dark matter cosmology with $H_0 = 70$ \kms\ Mpc$^{-1}$ and $\Omega_{\rm M} = 0.3$. In this cosmology, $1''$ corresponds to 5.4 kpc at the redshift of the quasar.

\section{ALMA Observations and Data Reduction}
\label{sec:obs}

Quasar P036$+$03 was observed with ALMA between 2021 August 2 and 2021 August 7 in five separate execution blocks. Minimum (maximum) antenna spacing varied between these observations from 47\,m (5.6\,km) on August 2 to 70\,m (8.3\,km) on August 7.  Four different amplitude calibrators were employed during the five different runs, but the same phase calibrator, quasar J0219$+$0120, was used for all observations. A second phase calibrator, quasar J0208$-$0047 was observed every 15 minutes in order to check the phase solution. Total on-source time on P036$+$03 was 3.3 hours in this extended configuration. The correlator was set up with one spectral window centered on the redshifted \CII\ emission line of P036$+$03 at 252.040 GHz using 480 3.9-MHz channels, while the remaining three spectral windows were set up to detect the continuum using the same spectral setup. We also use previous observations of P036$+$03, which were taken in a more compact configuration. These observations are described in \citet{Venemans2020}, and totaled 1.3 hours on-source.

We processed the data from the extended configuration using the ALMA pipeline \citep{Hunter2023}, which is part of the Common Astronomy Software Application package \citep[CASA v6.1.1-15;][]{McMullin2007,CASA2022}. After the initial processing, we performed minor additional manual flagging on one of the execution blocks. We then imaged both the extended and compact data together using the task \textit{tclean} within CASA. Because the extended and compact configuration were taken at the same phase reference and spectral setup, we did not perform any additional re-processing of the compact configuration data. The continuum was imaged using the multi-term, multi-frequency synthesis deconvolver with two Taylor coefficients and a reference frequency set to the frequency of the redshifted \CII\ line (252.040 GHz). We cleaned the image using the Briggs weighting scheme with a robust parameter of 0.5 down to twice the root-mean-square (rms) noise within a radius of 0\farcs5 of the center of the quasar. The rms noise of the final continuum image is 7.4 $\mu$Jy bm$^{-1}$, and has a beam size of 0\farcs073 by 0\farcs061 at a position angle of 88.6\degr. The \CII\ emission line was imaged in channels with a width of 30 \kms\ using the same weighting scheme and same clean region as the continuum image and was cleaned down to twice the rms noise per channel. This resulted in an rms noise of 96 $\mu$Jy bm$^{-1}$ per 30 \kms\ channel with a median beam size of 0\farcs083 by 0\farcs072 at a position angle of $-$71.4\degr.

\section{Analysis and Results}
\label{sec:analysis}

\begin{figure*}[!tbh]
\includegraphics[width=\textwidth]{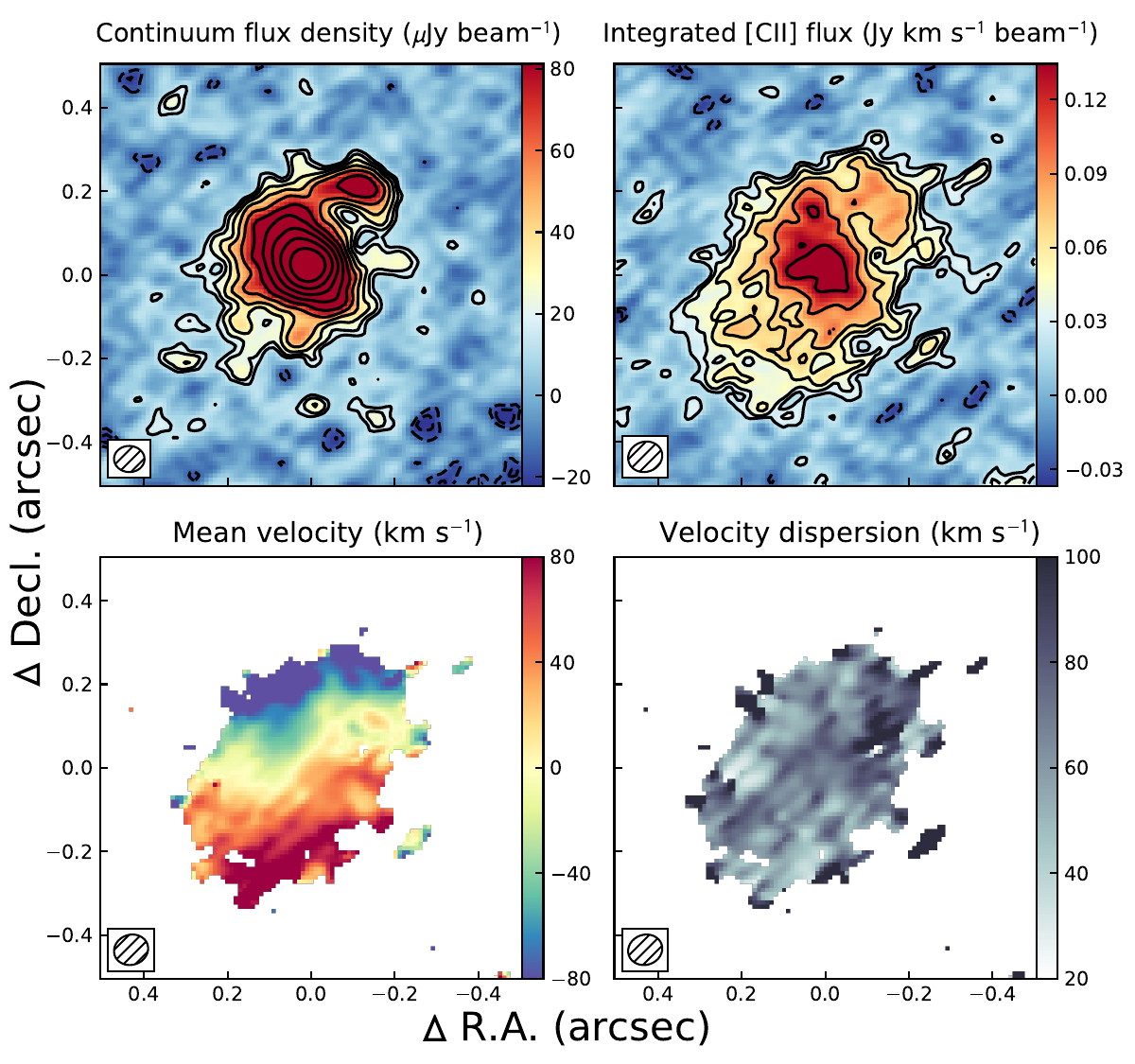}
\caption{Rest frame 1900\,GHz continuum (top left), integrated \CII\ flux density (top right), mean \CII\ velocity field (bottom left) and \CII\ velocity dispersion field (bottom right) of the quasar P036$+$03. In the top two panel the contours start at 2$\sigma$ and increase by powers of $\sqrt{2}$; negative contours at similar levels are dashed. The integrated \CII\ flux shows the emission over velocity range between -180 \kms\ and +180 \kms relative to $z = 6.5406$. The mean velocity field and velocity dispersion field were generated by fitting Gaussian profiles to the spectra at each pixel \citep[see further][]{Neeleman2021}. In all panels, the synthesized beam is shown in the bottom left corner.
\label{fig:contmom}}
\end{figure*}

\begin{figure*}[!tbh]
\includegraphics[width=\textwidth]{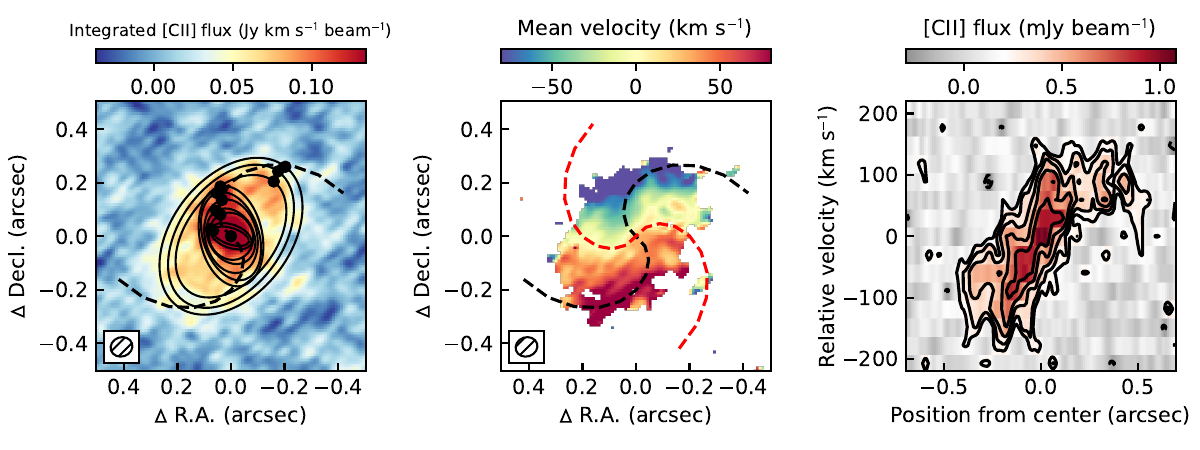}
\caption{\emph{Left panel:} integrated \CII\ emission of P036$+$03. The black ellipses are elliptical fits to the contours in the integrated \CII\ emission image, starting at 3$\sigma$ and increasing by $1\sigma$. The solid black circles mark the position of the apexes of the ellipses along the major axes. The dashed line is a simple parametric fit to these circles. \emph{Middle panel:} mean velocity field of the \CII\ emission line for P036$+$03 with the parametric fit overlayed. Also shown is the perpendicular curve to this fit (red dashed line). \emph{Right panel:} Position - velocity diagram along the parametric fitted line (black dashed line in left and middle panel). Positive x-values correspond to the southern `arm' of the parametric fit.
\label{fig:mom1pv}}
\end{figure*}

The new ALMA dust continuum images as well as maps of the integrated emission, mean velocity and velocity dispersion of the \CII\ emission line are shown in Fig.~\ref{fig:contmom}. The dust continuum centered at a rest-frame frequency of 1900\,GHz is resolved with a maximum extent of about 0$\farcs$4 (2.1 kpc) for the 3$\sigma$ contour. Besides strong, relatively compact emission from the central region of the galaxy, we also detect a highly significant (15$\sigma$) clump of dust continuum emission 0$\farcs$23 (1.3 kpc) northwest of the center. 

The integrated \CII\ flux density map, the mean velocity field and velocity dispersion field were created using the same method as described in \citet{Neeleman2021}. That is, for the integrated flux density maps we integrate all of the flux within the channels between $-$180 \kms\ and $+$180 \kms\ (see the channel map in Appendix \ref{sec:chanmap}). This velocity range maximized the integrated \CII\ emission \citep[see e.g.,][]{Venemans2020}. For the mean velocity field and velocity dispersion field, we fit a Gaussian emission line profile to the spectrum of each pixel. The mean velocity field and velocity dispersion field are the mean and the square root of the variance of this Gaussian profile, respectively. 

We find that the \CII\ emission has a larger extent and is less concentrated than the continuum, which is in agreement with other $z > 6$ quasar host galaxies \citep{Venemans2020}. These high resolution observations also confirm the conclusion from previous lower resolution observations that the emission arises from a single rotating disk galaxy and not a merging system  \citep{Neeleman2021}. At the new resolution, we resolve the \CII\ emission over many resolution elements, which shows that the \CII\ emission is smoothly varying with galactocentric radius, and the optical position of the quasar is in excellent agreement with the central position of this emission. Together with the relative constancy of the velocity dispersion across the \CII\ emission, this strongly disfavors a merger scenario. Fitting the \CII\ emission using a 2D Gaussian yields a full width at half maximum (FWHM) major axis of $0\farcs41 \pm 0\farcs03$ ($2.2 \pm 0.2$ kpc) at a position angle of $145 \pm 8\degr$. This corresponds to a maximum extent (diameter) of about $0\farcs7$ (3.8 kpc) for the 3$\sigma$ contour. The minor axis has a FWHM extent of $0\farcs29 \pm 0\farcs02$ ($1.6 \pm 0.1$ kpc), which results in an inclination estimate of $44 \pm 4$\degr\ if we assume a disk geometry (Section \ref{sec:kinematics}). The mean velocity field shows a velocity gradient across the galaxy consistent with ordered motion. However, the velocity field is distorted from the typical velocity field of a regularly rotating disk, which would show the velocity gradient aligned with the major axis of the emission at all radii (see section \ref{sec:kinematics}). For P036$+$03, the velocity gradient in the inner regions is instead aligned almost perpendicular to the outer major axis. Remarkably this distortion is also seen in the integrated emission map, where the high S/N contours are aligned with the minor axis, suggesting a common origin.

To illustrate this latter point, we fit ellipses to subsequent contours of S/N in the integrated \CII\ flux starting at 3$\sigma$ up to and including 12$\sigma$ in steps of 1$\sigma$. In Fig.~\ref{fig:mom1pv} we show these ellipses and mark the apexes of these ellipses with black points. The inner ellipses are clearly aligned with the minor axis, whereas the outer ellipses are aligned with the major axis of the galaxy. We also fit a simple parametric function of the form: $r(\theta) = a + m\theta$ to the apexes of the ellipses using a least-square optimization fitter. This fit is shown by the black dashed line in Fig. \ref{fig:mom1pv}. In the middle panel this fitted line accurately traces the steepest velocity gradient, whereas its perpendicular curve (red dashed line in Fig. \ref{fig:mom1pv}) traces the 0 \kms\ iso-velocity contour. This shows that the distortion in the velocity field is qualitatively similar to the distortion in the integrated \CII\ emission map. In the third panel of Fig. \ref{fig:mom1pv} we show the position-velocity ($p$-$v$) diagram along the black dashed line generated with the \textit{pvextractor} package \citep{PVextractor2016}. The $p$-$v$ diagram shows a rising velocity curve that starts to flatten at a radius of $\sim$$0\farcs1$ (540 pc) to a maximum rotational velocity not corrected for inclination of $100 \pm 10$ \kms.

Despite this distorted velocity field, the velocity dispersion across the disk is relatively small with a median velocity dispersion of 66 \kms, which is much less than the average velocity dispersion among $z \sim 6$ quasar host galaxies of 129 \kms\ \citep{Neeleman2021}. Using the approach as described in \citet{Walter2022}, we can estimate the scale height of the disk from this velocity dispersion and the gas mass. The latter was estimated from the far-infrared dust continuum measurement in \citet{Neeleman2021}. The scale height, $h_{z}$, is then given by $h_z = \kappa r_{\rm maj}^2 \sigma_v^2 / 2 G M_{\rm gas}$, where $r_{\rm maj}$ is the major axis extent of the galaxy and $\kappa$ is a scale factor of order unity that depends on the assumed functional form of thickness of the disk (for an exponential function $\kappa = 2$ and for a Gaussian function $\kappa = \sqrt{\pi}$). For P036$+$03, the corresponding scale height estimate is $80 \pm 30$ pc, where the uncertainty includes both the observational uncertainty and the systematic uncertainty of the assumed thickness profile. This scale height is much smaller than the disk scale length ($h_R \equiv {\rm FWHM} / 2\sqrt{2 \ln 2} = 0.94$ kpc), with a scale height to disk scale length ratio of about 12. This suggests that the gas in the host galaxy of P036$+$03 is constrained to a very thin, kinematically disturbed disk. In Section \ref{sec:barorwarp} we explore the potential causes for the distortion in both the integrated \CII\ emission and the velocity field.

\subsection{Dust vs [CII] comparison}
\label{sec:ciidef}

\begin{figure}[!hbt]
\includegraphics[width=0.48\textwidth]{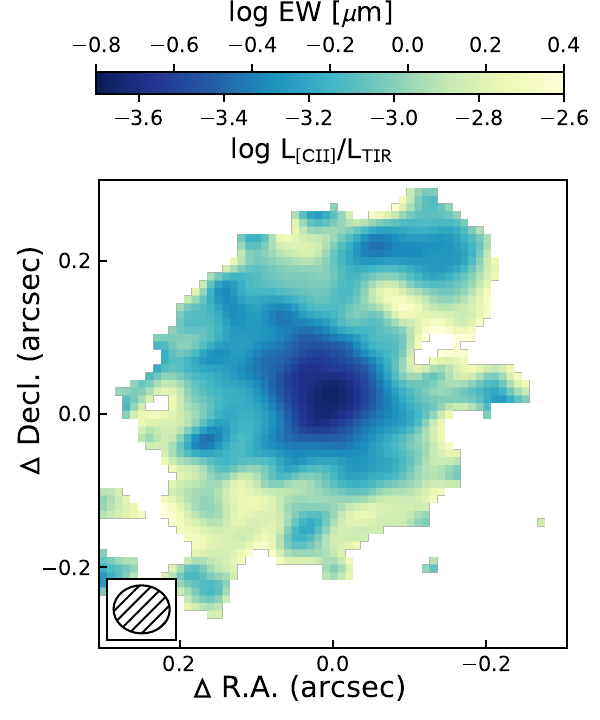}
\caption{Spatially resolved \CII\ deficit for P036$+$03. Only pixels are shown for which both the 1900 GHz dust continuum emission and the \CII\ emission line is detected at $>$2$\sigma$. The color bar shows the ratio of these two measurements both expressed as an equivalent width and as a luminosity ratio. However, we caution the reader about the conversion factors used in obtaining this latter expression \citep[see][]{Walter2022}.  The average synthesized beam is shown in the bottom left inset. 
\label{fig:ciidef}}
\end{figure}

As was mentioned in the previous section, the dust continuum measurement for P036$+$03 is much more centrally concentrated than the \CII\ emission. This can best be seen by plotting the equivalent width of the \CII\ line (EW$_{\rm [CII]}$), which is proportional to the ratio of these two ALMA measurements:  i.e., ${\rm EW_{[CII]}} = \lambda_{\rm [CII]} / c \int{S_{\rm [CII]}}dv / S_{\rm 1900\, GHz}$ (Fig. \ref{fig:ciidef}). The equivalent width strongly decreases towards the center of the galaxy, which is in agreement with the bulk of $z > 6$ quasar host galaxies and local ultra luminous infrared galaxies \citep{Venemans2020,Herrera-Camus2018}. 

We also converted the equivalent width measurement to a \CII\ luminosity over total far-infrared luminosity ratio ($L_{\rm [CII]} / L_{\rm TIR}$) assuming a canonical modified black body \citep[$T_d = 47$\,K and $\beta = 1.6$; see e.g.,][]{Beelen2006}. However, we caution about using this approach, because the dust temperature (and related far-infrared dust continuum field) will likely be higher at the center, which will strongly affect the calculated total infrared luminosity from this single point dust continuum observation \citep[see][]{Walter2022}.

What is interesting to note in Fig. \ref{fig:ciidef} is that the clump seen in the continuum $0\farcs23$ northwest of the center of the galaxy has a significantly lower EW$_{\rm [CII]}$ than the parts of the galaxy at a similar galactocentric radius. We like to stress that this lower EW$_{\rm [CII]}$ is driven by an increase in the dust continuum emission because we also see a slight excess of \CII\ emission within the clump (Fig. \ref{fig:contmom}). One possible interpretation of this lower EW$_{\rm [CII]}$ is that the dust temperature and far-infrared radiation field are higher in this region due to on-going, intense star-formation. This is also observed in star-forming regions of local galaxies \citep{Smith2017}. A Gaussian fit to the continuum image of this region yields an upper limit on the deconvolved size of the region of $60 \pm 6$ parsec, although this size estimate is strongly dependent on how much of the dust emission is assigned to the clump instead of the underlying extended dust continuum emission from the disk. However, even if we conservatively take a diameter of 400 parsec (the size of the beam) as the upper limit to the size of the region in the following estimations, then the size remains well within the expected range of large star-forming clumps in local galaxies \citep[e.g.,][]{Elmegreen1996}. To calculate the SFR surface density of this region, we can convert the total dust continuum flux arising from this clump into a SFR using the scaling relation of \citet{Kennicutt2012}. Here we assume that all of the emission is due to SFR and we further assume that the far infrared emission satisfies the same canonical modified black body as described in the previous paragraph. This yields a SFR of $150\pm20\ M_\odot$ yr$^{-1}$ and thus a minimum star-formation surface density of $>$$1 \times 10^3\ M_\odot$ yr$^{-1}$ kpc$^{-2}$. These observations all support that this region is a large star-forming clump within the disk of the galaxy. In Section \ref{sec:clump} we speculate on the origin of this star-forming region.

\subsection{Kinematic Modeling of CII}
\label{sec:kinematics}

\begin{figure*}[!tbh]
\includegraphics[width=\textwidth]{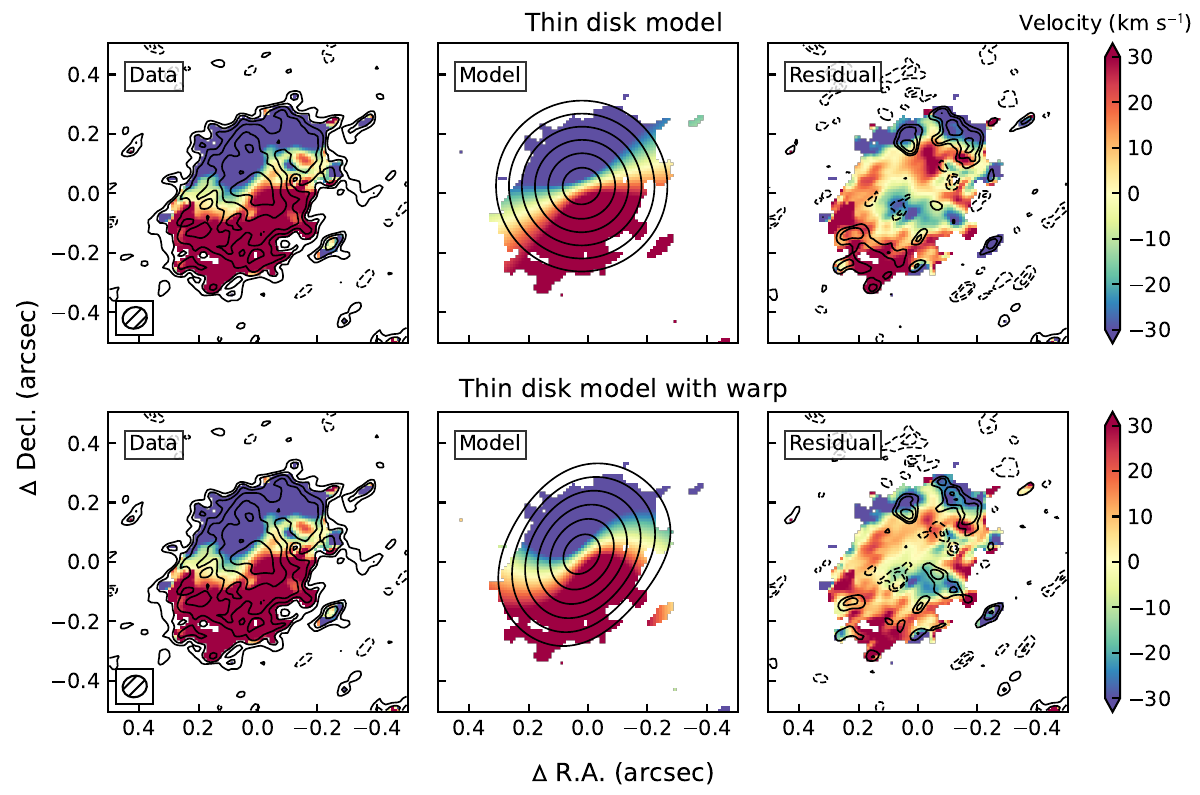}
\caption{Kinematic modeling of the \CII\ emission line for P036$+$03. The top row shows the results from a simple thin disk model, whereas the bottom row shows the results of a thin disk model that has a warp. The left panel of each row shows the data, the middle panel the model and the right panel the residual. In all cases the raster image is the velocity field whereas the contours show the integrated \CII\ emission line. The plotted range in the velocity field has been truncated to $\pm$30 \kms\ in order to highlight the distortion in the velocity field and the residuals in the rightmost panels. Contours start at 2$\sigma$ and increase by powers of $\sqrt{2}$; negative contours at similar levels are dashed. The bottom left inset in the left panels shows the synthesized beam of the observations. 
\label{fig:kinmodel}}
\end{figure*}

The kinematics of the \CII\ emission line for P036$+$03 was discussed previously in \citet{Neeleman2021}. Based on the $0\farcs14$ resolution data, P036$+$03 was classified in this paper as a disk galaxy. However, \citet{Neeleman2021} reported that P036$+$03 had the largest deviation from the simple disk model used in their analysis out of all the galaxies in the sample. This was attributed to a non-azimuthal \CII\ emission distribution either caused by a warp in the disk or due to the presence of spiral arms. The higher resolution observations discussed here confirm this assertion and indeed show a clear distortion in both the velocity field as well as non-azimuthal \CII\ emission in the integrated \CII\ line flux density.

For comparison, we have re-run the kinematic modeling performed in \citet{Neeleman2021} on the higher resolution data using the same simple thin disk model as presented in \citet{Neeleman2021}. The kinematic fitting was done using the \textit{qubefit} package. In short, \textit{qubefit} fits a user-defined model to the data cube, and applies a Bayesian Markov Chain Monte Carlo approach to estimate the posterior distribution of each of the parameters in the model. The advantage of \textit{qubefit} over other fitting codes is the limited number of free parameters (9 for the thin disk model), and its ability to easily integrate custom models. For more details of the fitting code, we refer the reader to \citet{Neeleman2021}.

The results of the thin disk model are shown in Figure \ref{fig:kinmodel}, which shows that the simple disk model cannot accurately reproduce the \CII\ emission. The disk model is roughly face-on, which results in positive residuals in both the northwest and southeast part of the galaxy where there is excess \CII\ emission in the data compared to the model. For the velocity field residuals, we see that there remains a distinct velocity gradient in the center part of the galaxy where the position angle of the model is not aligned with the direction of the maximum change in the velocity field. To estimate the goodness-of-fit of the model, we calculate the reduced-$\chi^2$ and Kolmogorov-Smirnov (KS) one sample test probability for a mask that is the union of the 2$\sigma$ data and model contours sampled at the size of the beam \citep[see][for an explanation of this choice]{Neeleman2021}.  For this model, we find a reduced-$\chi^2$ value of 1.92 (Bayesian Inference criterion, BIC, of 835) and a KS probability of 0.38\% that the residuals within the mask are consistent with the Gaussian noise distribution of the data set.

To account for the non-azimuthal emission in P036$+$03, we have modeled this data set using a slight adaption of the thin disk model. In this model, we allow the position angle of the galaxy to vary linearly with galactocentric radius. That is, we assume the position angle of the galaxy (PA) has the form: ${\rm PA}(r) = {\rm PA}_0 + {\rm PA}_1 r$. This adaptation introduces a warp to the galaxy, whereby the inner and outer regions are not in the same plane. This warped disk model adds a single parameter to the thin disk model, and the results of the fitting procedure are shown in the bottom row of Fig. \ref{fig:kinmodel}. 

As Fig. \ref{fig:kinmodel} shows, this warped disk model recovers the actual observations better. We can see that the integrated \CII\ emission of the model matches the overall shape of the observed integrated \CII\ emission. This model can account for most of the emission in the southeast corner of the galaxy, while reducing the residual flux in the northwest quadrant. What is particularly noteworthy is that the model recovers the distortion seen in the velocity field (bottom middle panel). This reduces the structure in the residual velocity field both in the inner part of the galaxy as well as the outer parts. The reduced-$\chi^2$ and KS probability are 1.73 (BIC of 763) and 1.73\% respectively. These values indicate that despite being a better fit to the data compared to the thin disk model, the warp disk model remains unable to recover all of the structure in the \CII\ emission. This discrepancy is largely driven by the small scale structure that exists within the \CII-disk. These structures cannot be reproduced with the adopted azimuthally symmetric model. This can be further seen in the residual channel maps that are shown in Appendix \ref{sec:modelfit}.

We note that this simple warped disk model is not a fully generalized warped disk model, since it assumes a constant inclination across the galaxy. A generalized warped disk will also vary in inclination with radius. Therefore,  we have also fitted the data with a more generalized warped disk model in which we vary the inclination linearly with radius. However, this model fits the data equally well as the model with only a varying PA. This indicates that the inclination is roughly constant, which is corroborated by the elliptical fits of the data at varying S/N in Section \ref{sec:analysis}). We therefore opt to use the model with less parameters (and thus a lower BIC score) in the below analysis. Finally, we have also fitted P036$+$03 using the fitting code $^{\rm 3D}$Barolo \citep{DiTeodoro2015}. This tilted-ring fitting code also recovers the warp in the disk, and gives consistent values for both the position angle, inclination and the rotation velocity albeit using a much larger number of free parameters. We again opt to use the model with the least free parameters.

For the warped disk model, we find that the position angle varies from $(226 \pm 3)$\degr\ in the center to $(164 \pm 5)$\degr\ in the outskirts of the galaxy with a constant inclination of $(40.4 \pm 1.3)$\degr. This inclination estimate is consistent with the inclination estimate obtained from comparing the major and minor axis (Section \ref{sec:analysis}). The best-fit model has a constant, inclination-corrected rotational velocity of $(116 \pm 3)$ \kms\ and constant velocity dispersion of $(66.4 \pm 1.0)$ \kms. This maximum rotational velocity is significantly smaller than the maximum rotational velocity found in the low resolution data of P036$+$03 \citep{Neeleman2021} due to the increased inclination estimate for the galaxy. Together with the updated extent from these high resolution observations, we determine a new dynamical mass estimate of $M_{\rm dyn} = (1.96 \pm 0.10) \times 10^{10} M_\odot$. This new dynamical mass estimate has propagated observational uncertainties of less than 5\%, and thus the overall uncertainties are driven by the systematic uncertainties resulting from the chosen model, which are likely dominated by the exact parametric description of the warp. We note, however, that this dynamical mass estimate and the model-agnostic \CII\ line width methods ($M_{\rm dyn} = (1.5 \pm 0.9) \times 10^{10} M_\odot$) described in detail in \citet{Neeleman2021} are in good agreement. The lower dynamical mass estimate only strengthens the conclusion in \citet{Neeleman2021} that the galaxy hosting the supermassive black hole in P036$+$03 is under-massive compared to the mass of the black hole derived from the \ion{Mg}{2} line \citep[$M_\bullet = (3.7 \pm 0.6) \times 10^9 M_\odot$;][]{Farina2022}.

\section{Discussion}
\label{sec:discussion}
 
In this section we discuss the results from the analysis described in Section \ref{sec:analysis}. In particular, we discuss possible causes for the disturbed kinematics of P036$+$03 and possible effects that this disturbed kinematics has on the evolution of the galaxy.

\subsection{Bar or warped disk?}
\label{sec:barorwarp}
Several competing scenarios could explain the observed kinematic distortion; we will discuss the two most likely scenarios here, which is that the kinematic distortion is due to a bar structure within the disk or that the kinematic distortion is due to a warp in the disk.

The first scenario, the existence of a bar within the disk of P036$+$03, is scientifically intriguing as it provides a mechanical way for gas to lose angular momentum and migrate toward the center of the galaxy \citep[e.g.,][]{Shlosman1990}. A fraction of this low angular momentum gas could, in turn, get accreted onto the supermassive black hole, providing a way to power the central luminous quasar. Bar-driven active galactic nuclei (AGN) are observed at low redshift albeit for SMBHs that are much less massive \citep[$M_\bullet \lesssim 10^8 M_\odot$;][]{Carollo2002,Silva-Lima2022}.

Observations of local barred galaxies have revealed that bars can contort the velocity field due to radial motions along the bar \citep[e.g.,][]{Gadotti2020}. In general this would contort the iso-velocity contours into a characteristic integral sign shape similar to the velocity field shown in Figure \ref{fig:contmom}. Together with the observed elongation of the \CII\ and dust continuum emission along the minor axis of the galaxy, which in this scenario would be due to excess star-formation within the bar, these two features together support the bar scenario.

There are, however, several features that run counter to the expectation of a bar structure. First is that at low redshift, bar instabilities predominantly power low-luminosity AGNs, not extremely luminous quasars like P036$+$03 \citep[e.g.,][]{Treister2012}. Unless the physical processes that power bar-driven AGNs at high redshift produce significantly more luminous quasars, the occurrence of the bar and the luminous quasar in P036$+$03 would need to be coincidental. However, more difficult to explain with the bar scenario is that bar-instabilities distort the velocity field in the inner galaxy to a radial extent which is approximately the size of the bar. The velocity field of the outer galaxy would continue to resemble that of a standard rotating disk. This is contrary to the velocity field of P036$+$03, which shows a disturbed velocity field across the galaxy. Even at the outermost (3$\sigma$) contour, the velocity gradient does not point along the major axis, which is indicative of a continued torque at these radii. 

We stress that neither of these features definitively rule out a bar structure for P036$+$03. A large bar structure or tri-axial bulge that would extend out to the radius probed with the \CII\ emission would explain the observed velocity field. Furthermore, $z = 6.5$ galaxies are known to be much more gas-rich than low redshift counterparts, \citep{Carilli2013,Tacconi2020}, which could provide the necessary fuel to power the luminous quasar. 

The second scenario is that the kinematic distortion is driven by a warp in the disk. This could be due to a recent merger or other external torques such as gas inflows. The velocity field of P036$+$03 is fully consistent with the expected velocity field of a warped disk as seen in local warped galaxies \citep[e.g.,][]{Jozsa2007} and the velocity field from a simple warped disk model described in Section \ref{sec:kinematics}. By comparing the position angle of the major axis in the inner and outer disk, we conclude that the warp needs to be at least 60$\degr$, which is well within the range observed in local warped disk galaxies \citep{Jozsa2007}. 

There are two advantages of this scenario compared to the bar scenario. First, it can easily explain the warped velocity field across the full disk, as outside torques are expected to influence the kinematics of the full disk, not just the inner region. Second, it fits in with theoretical expectations that copious amounts of gas is needed to grow and evolve massive quasar host galaxies \citep[e.g.,][]{Inayoshi2020}. We, therefore, favor this interpretation. We note that about one third of all $z>6$ quasar host galaxy shows evidence of disk rotation \citep{Neeleman2021} and due to the expected ubiquity of gas flows into these galaxies, we hypothesize that many of these disks are warped. However, higher resolution observations of these disks are needed to confirm this hypothesis. In the next section, we describe observational evidence for the possible cause of this outside torque.

\vspace{1cm}

\subsection{Causes for a warped disk}
\label{sec:gasflows}

\begin{figure}[!t]
\includegraphics[width=0.47\textwidth]{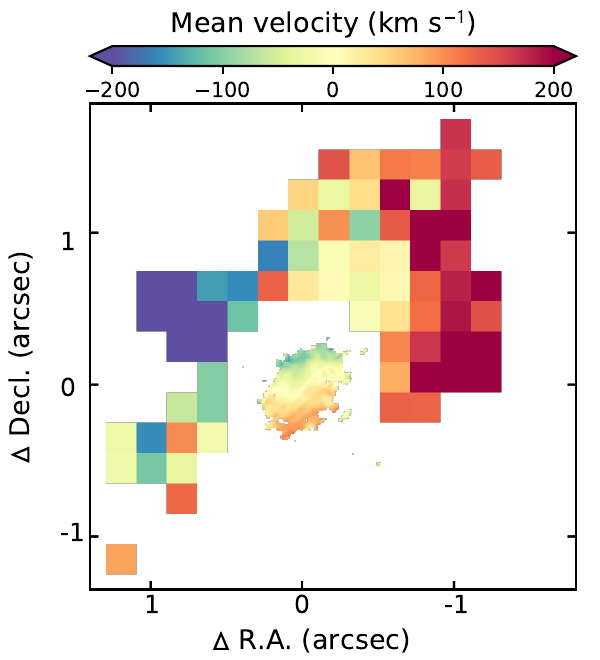}
\caption{Velocity field for the interstellar medium and circumgalactic medium of P036$+$03 as traced by \CII\ and Ly$\alpha$, respectively. The velocity field of the \CII\ emission line is the same as Figure \ref{fig:contmom} albeit at a different velocity scale. The velocity field of the \lya\ line is described in detail in \citet{Drake2022} and is shown here at the native pixel scale of the VLT/MUSE observations ($0\farcs2$). For the MUSE observations, only \lya\ emission belonging to the halo is shown \citep[see][]{Farina2019}. Unlike the \CII\ velocity field, the \lya\ velocity field is the flux-weighted mean velocity, which is the first moment of the data cube along the spectral direction. However this different approach does not affect the conclusions in this paper. 
\label{fig:almamuse}}
\end{figure}

Having established that the disk likely contains a warp, we can assess the causes for this warp. In general, warps are predominantly caused by external forces, in particular mergers and interactions between the galaxy's interstellar medium and the halo \citep[e.g.,][]{Jozsa2007}. For $z > 6$ quasar host galaxies, mergers are a particularly attractive solution because simulations suggest that mergers play a dominant role in the formation of luminous quasars at high redshifts \citep[e.g.,][]{Hopkins2008}, and we observe that about a third of all luminous, $z > 6$ quasars are actively undergoing interactions or have a nearby close companion \citep[e.g.,][]{Decarli2017,Neeleman2019,Neeleman2021}. However, P036$+$03 shows no evidence of a recent major merger or nearby companion in either its dust continuum emission or \CII\ emission. Indeed the velocity-integrated \CII\ line flux density is one of the most regular, elliptical disks in the full sample of quasars targeted in \citet{Neeleman2021}. In addition, interactions with companion galaxies will often produce asymmetric warps \citep{Weinberg2006}, while the warp in P036$+$03 is remarkably symmetric. Finally estimates of the quasar lifetime from the quasar's proximity zone \citep[][]{Eilers2017}, suggests continuous quasar activity for at least $10^6$ years, which are consistent with a  quiescent merger-free history. The observations therefore disfavor a major merger as the origin of the warp.

Alternatively, the warp could be caused by torques between the galaxy's halo and the interstellar medium. Although these warps have a tendency to dissipate quickly, theoretical studies have shown that external flows of gas onto the halo can sustain a warp \citep{Ostriker1989}. Cosmological simulations have shown that gas flows streaming in from the cosmic web are expected to be most dominant at high redshift \citep[e.g.,][]{Keres2005,Dekel2009}, and provide the necessary fuel to sustain the growth galaxies \citep[e.g.,][]{Walter2020}. These gas flows are therefore expected to play a crucial role in the formation of high redshift galaxies and the built up of the extended halo gas, i.e., the circumgalactic medium. \citet{Drake2022} showed that the halo kinematics is decoupled from the interstellar medium. Therefore as gas gets deposited onto the galaxy from the halo, the difference in angular momentum will exert a torque on the disk \citep[see e.g.,][for simulations of this process]{Danovich2015} that could warp the disk.

In Figure \ref{fig:almamuse} we show the mean velocity field for both the \CII\ emission and the \lya\ emission for P036$+$03. Despite the difficulty with assessing kinematics from the scattered \lya\ line \citep[e.g.,][]{Dijkstra2012}, the MUSE observations show a clear velocity gradient from east to west. This velocity gradient is in excess of 400 \kms, significantly larger than the uncertainty on the observations, which is less than 40 \kms. As was noted in \citet{Drake2022}, this velocity gradient is misaligned with the velocity gradient of the \CII\ emission line. With the higher resolution \CII\ data, we can constrain the misalignment to be $120 \pm 30$\degr\ between the \lya\ velocity gradient and the \CII\ velocity gradient at the largest observed radius. If we assume that gas is being deposited from the circumgalactic medium onto the interstellar medium, then the angular momentum of this gas would be misaligned with the angular momentum of the disk and provide the external torque needed to warp the disk. The observations of P036$+$03 are fully consistent with this scenario. Even the relatively low velocity dispersions across the disk are consistent with this scenario as warp flaring or disk heating, which are two common ways to increase velocity dispersions within a disk, are not linked to warps \citep{GarciadelaCruz2023}. This is likely because warps result from events that produce only modest gravitational disturbances, unlike major mergers. We note that this scenario also encompasses minor mergers, as minor mergers are in many ways indistinguishable from accretion of a clumpy gas flow \citep{Dekel2009}. 

Despite the agreement between this scenario and the observations, it remains difficult to quantitatively assess the kinematics between the circumgalactic medium and the disk. This is partly due to the previously mentioned difficulty with assessing the kinematics of \lya\ and partly because we lack a tracer that connects the kinematics between the circumgalactic medium and the disk. As a result, it is hard to quantify the full three-dimensional spatial distribution of the halo gas, which is necessary to accurately model the gas kinematics of the halo. Nevertheless, qualitatively the observations agree and we therefore favor this scenario over major mergers.

\subsection{Warp-induced star-formation}
\label{sec:clump}
Besides looking at possible causes for the warp, we can also explore how the warp is affecting the properties of the host galaxy; in particular its ability to form stars. As was noted in Section \ref{sec:ciidef}, there is excess continuum emission 0$\farcs$23 (1.3 kpc) northwest of the galaxy, which is likely a large star-forming region. One possibility is that this region is the stellar remnant of a recent minor merger. However, it is interesting to note that this star-formation region falls right on the parametric fit to the apexes of the ellipses describing the warp (see Section \ref{sec:analysis} and the black dashed line in Fig. \ref{fig:mom1pv}). It is in this region of the warped disk that adjacent orbits have their closest approach; providing a way for gas to collide, exchange angular momentum and thus form stars. We note that this is also seen in local warped disk galaxies, where spiral arms often trace this `line of apexes' \citep{Reshetnikov1998}. In this latter scenario the warp would be directly responsible for the formation of stars.

We can get an estimate of the stability of this star-forming region against gravitational collapse by measuring the Toomre-Q parameter \citep{Toomre1964,Goldreich1965}. By assuming the far-infrared dust continuum emission satisfies a modified black body curve ($T_{\rm dust} = 47$ K, $\beta = 1.6$) and a dust-to-gas ratio of 100, we get a gas surface density of $1.5 \times 10^{4} M_\odot$ pc$^{-2}$ \citep[see e.g.,][]{Venemans2012} within the 400 pc region. With a velocity dispersion of 66 \kms, rotational velocity of 116 \kms, and radial distance of 1.3 kpc, this yields a Toomre-Q parameter of about $\sqrt{2}\sigma_v v_{\rm rot} / \pi G r \Sigma_{\rm gas} = 0.04$ for this region. Even without considering other forms of baryons, which would lower the Q-parameter even further \citep[see e.g.,][]{Neeleman2019}, this value is much smaller than one, and therefore the gas is gravitationally unstable. This suggests that conditions within this region are suitable for gas to collapse and form stars. This provides further evidence that the excess dust continuum emission within this region is driven by star formation, and its position within the galaxy is consistent that this is driven by the warp in the disk.

\section{Summary and Conclusions}
\label{sec:conclusions}

We have presented ALMA $0\farcs075$ ($\approx$400 pc) resolution dust continuum and \CII\ line emission imaging of the $z=6.5406$ quasar host galaxy P036$+$03. Previous lower resolution imaging revealed that this galaxy showed evidence of disk rotation. The new higher resolution observations confirm that the \CII\ emission arises from a thin gaseous disk. Rough estimates of the scale height of the galaxy suggests the disk has a scale length to scale height ratio of about 12. With over 50 resolution elements across the disk traced by \CII, we can provide a detailed description of the disk kinematics. We find that the \CII\ velocity field is distorted from a regular rotating disk, whereas the velocity dispersion field remains relatively quiescent and constant with a mean value of 66 \kms.

Although both internal (e.g., bars) as well as external torques (e.g., mergers or gas inflows) can create a distorted velocity field, we assert that external torques are the cause for the velocity distortion seen in P036$+$03. Observations of the \lya\ line with VLT/MUSE reveal that gas within the halo of P036$+$03 appears to be counterrotating compared to the disk of the galaxy. This suggests that interactions between the halo gas and the gas within the disk of the galaxy might be responsible for the torque needed to induce a warp in the disk. About one third of all $z>6$ quasar host galaxy shows evidence of disk rotation \citep{Neeleman2021} and due to the expected ubiquity of external torques, we hypothesize that many of these disks are likely warped. However, higher resolution observations are needed to confirm this hypothesis. The fact that such a high fraction of quasar host galaxies show regular rotation is a testament to the resistance and/or recovery speed of these disks to outside torques.

The high resolution observations also revealed a dust continuum emitting region $0\farcs23$ (1.3 kpc) offset from the center of the galaxy. We posit that this region is a star-forming region that was formed due to the warp in the disk, because the star-forming region is located within the region of the warped disk where gas can easily collide and lose angular momentum. If our hypothesis is correct, the combined MUSE and ALMA observations reveal a cohesive picture how gas flows from the halo can trigger star formation. To be specific, within the disk-halo interface, the mismatch between the angular momentum of the halo gas and the gas within the disk causes the disk to warp. This warp, in turn, provides regions within the disk where gas can collapse and form stars. Although this hypothesis can explain the observations, a larger sample is needed to assess its validity.

These observations highlight the great synergy that exists between optical and far-infrared observations. They also demonstrate the feasibility of obtaining high resolution, high sensitivity observations of the interstellar medium of $z > 6$ quasar host galaxies. With a larger sample we could provide a statistical view of the processes that drive the growth of these massive galaxies. Finally it is worth noting that the James Webb Space Telescope is poised to provide a completely new light on the black hole properties of these quasars, and through observations of nebular emission lines such as H$\alpha$ and [\ion{O}{3}], can provide a better, non-scattered, estimate of the kinematics of the ionized gas surrounding these galaxies. Together, these state-of-the-art observatories will allow for a comprehensive picture of the first billion years of growth of the most massive galaxies in the universe.

\begin{acknowledgments}
We would like to thank the anonymous referee for their comments and suggestions, which helped improve the manuscript.
This paper makes use of the following ALMA data: ADS/JAO.ALMA\#2019.1.01633.S. ALMA is a partnership of ESO (representing its member states), NSF (USA) and NINS (Japan), together with NRC (Canada), MOST and ASIAA (Taiwan), and KASI (Republic of Korea), in cooperation with the Republic of Chile. The Joint ALMA Observatory is operated by ESO, AUI/NRAO and NAOJ. The National Radio Astronomy Observatory is a facility of the National Science Foundation operated under cooperative agreement by Associated Universities, Inc. M.N., F.W. and R.A.M. acknowledge support from ERC Advanced grant 740246 (Cosmic\texttt{\char`_}Gas).
\end{acknowledgments}

\facilities{ALMA, VLT/MUSE}
\software{Astropy \citep{Astropy2013,Astropy2018}, CASA \citep{CASA2022},  Matplotlib \citep{Hunter2007}, Numpy \citep{Harris2020}, PVextractor \citep{PVextractor2016}, Qubefit \citep{Neeleman2021}}

\bibliography{bib}
\bibliographystyle{aasjournal}

\appendix
\section{Channel Map of P036+03}
\label{sec:chanmap}
\begin{figure*}[!htb]
\centering
\includegraphics[width=\textwidth]{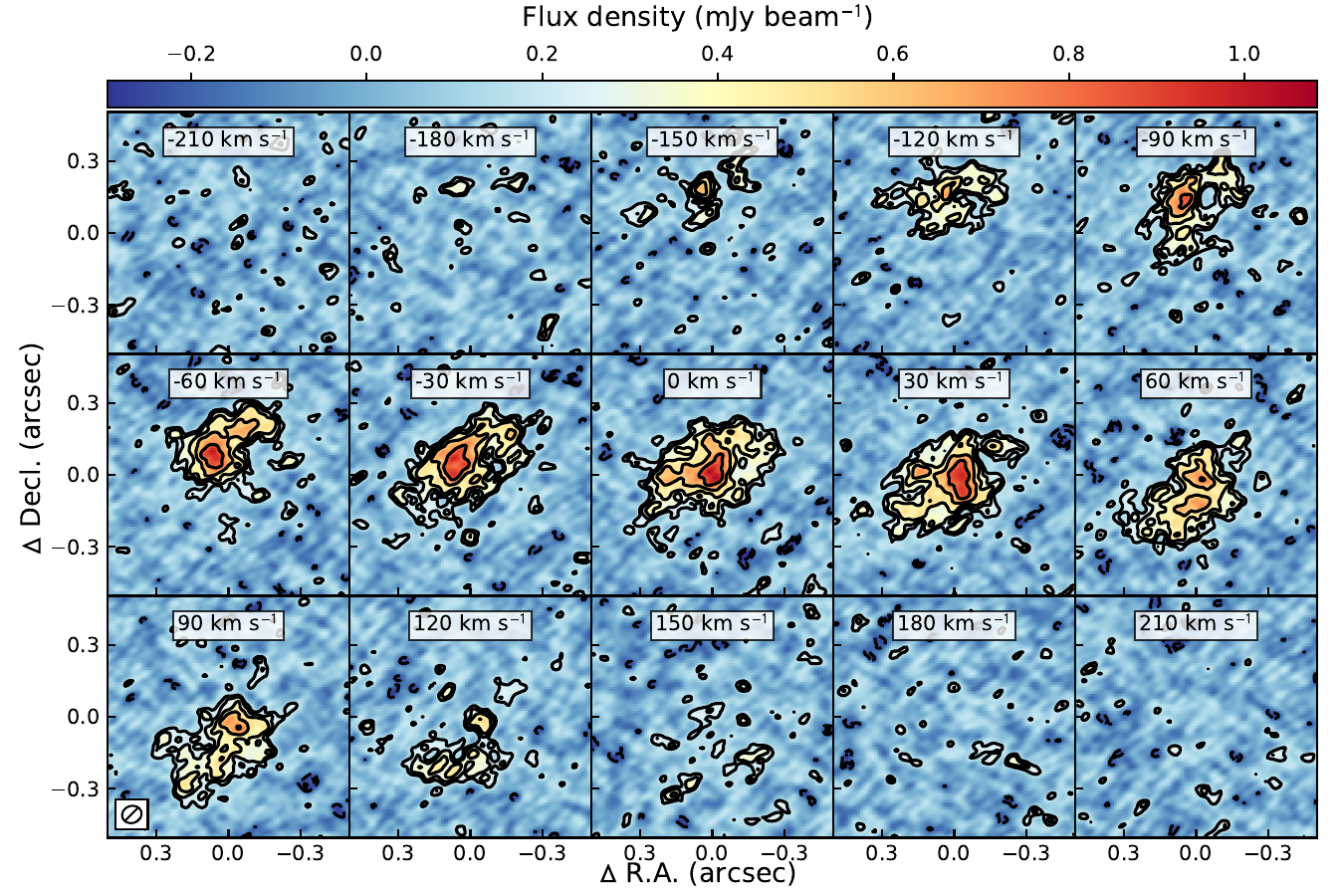}
\caption{Channel maps of the \CII\ emission line for P036$+$03. The redshift reference for the relative velocities is $z=6.5406$ (252.040 GHz). Contours start at 2$\sigma$ and increase in powers of $\sqrt{2}$, where $\sigma = 96\ \mu$Jy\,bm$^{-1}$. Negative contours at similar levels are shown in the figure with a dashed line. In the bottom left panel we show the size of the synthesized beam for the observations. 
\label{fig:cm}}
\end{figure*}

Figure \ref{fig:cm} shows the channel maps of the \CII\ emission line for P036$+$03.

\section{Model fitting of P036+03}
\label{sec:modelfit}
\begin{figure*}[!htb]
\centering
\includegraphics[width=\textwidth]{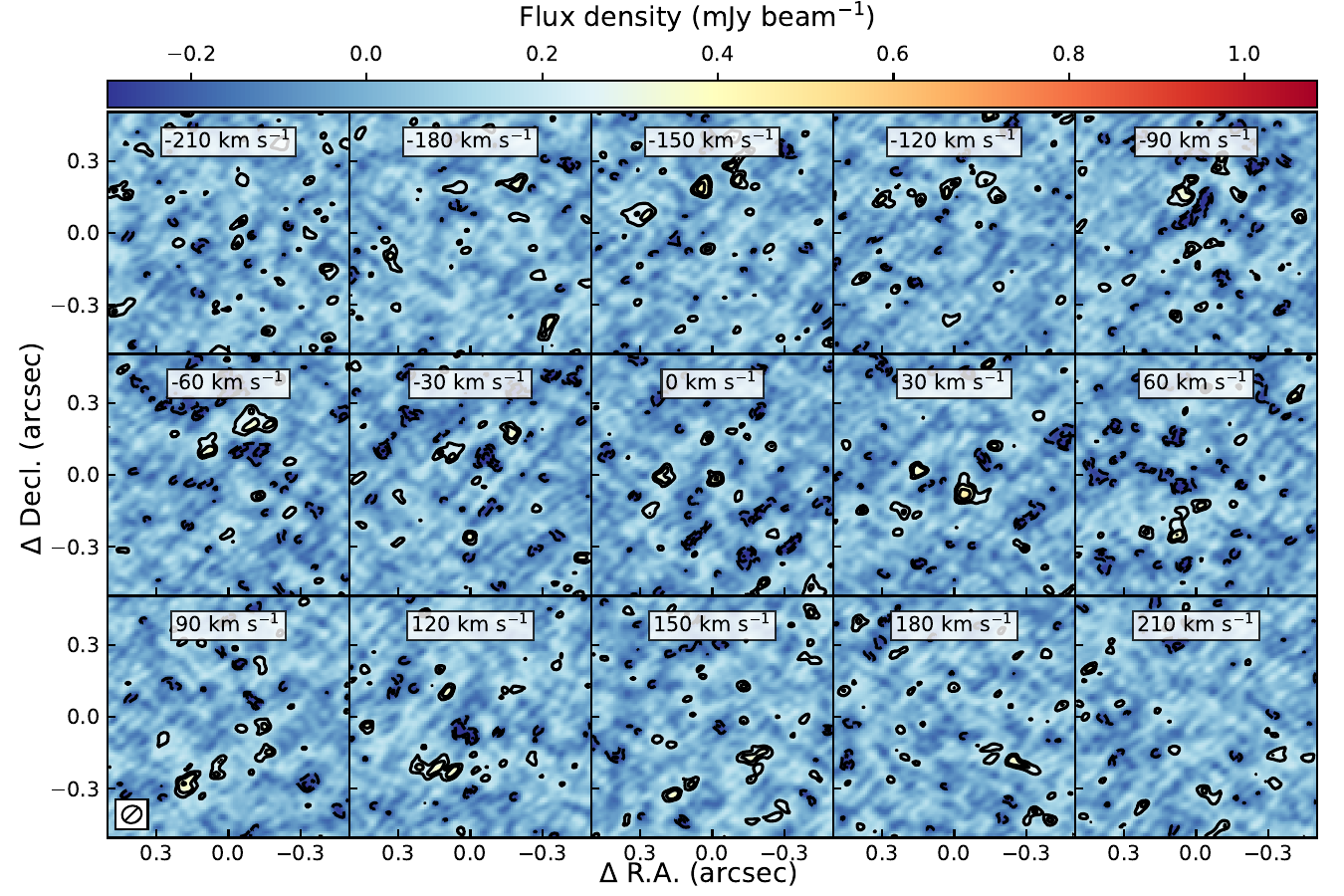}
\caption{Channel maps of the residual after subtracting the best-fit warped disk model. The figure has the same scale, and contours are drawn at the same levels as in Fig. \ref{fig:cm}.  
\label{fig:cmres}}
\end{figure*}

Figure \ref{fig:cmres} shows the residual channel maps of the \CII\ emission line for P036$+$03 after subtracting the best-fit warped disk model described in the text. The contours are drawn at the same levels as in Fig. \ref{fig:cm}. The majority of the large scale \CII\ emission can be accounted for with this azimuthally averaged model, but there remain small regions of both excess and deficient emission on small scale that are inconsistent with the model. These regions are likely real clumpy substructure within the \CII\ emission. The parameters for this fit are listed in Table \ref{tab:diskfit}, as well as the best-fit parameters for the disk model without the warp, which is the same as the model run in \citet{Neeleman2021} except on the higher resolution data. The uncertainties on the model parameters are the 1$\sigma$ (i.e., 16$^{\rm th}$ and 84$^{\rm th}$) percentile of the posterior distribution of the parameters. For comparison, we can estimate the uncertainty on the velocity-integrated flux density because this quantity is proportional to ($I_0 R_{\rm D}^2 \sigma_v$), which yield a total velocity-integrated flux density of $3.32 \pm 0.11$ Jy \kms. This is in excellent agreement with the observed velocity-integrated flux density of $3.32 \pm 0.10$ Jy \kms, which in turn is in good agreement with the flux density from the lower resolution data alone \citep[see][]{Venemans2020}.

\begin{deluxetable*}{lllllllllll}[!tb]
\tabletypesize{\scriptsize}
\tablecaption{Parameters of the kinematic modeling with a disk model
\label{tab:diskfit}}
\tablehead{
\colhead{Model} &
\colhead{R.A. ($x_c$)} &
\colhead{Decl. ($y_c$)} &
\colhead{$z_{\rm kin}$} &
\colhead{P.A.$_{0}$} &
\colhead{P.A.$_{1}$} &
\colhead{$i$} &
\colhead{$v_{\rm rot}$} &
\colhead{$\sigma_v$} &
\colhead{$I_0$} &
\colhead{$R_{\rm D}$} \\
\colhead{} &
\colhead{(ICRS)} &
\colhead{(ICRS)} &
\colhead{} &
\colhead{($\degr$)} &
\colhead{($\degr$ kpc$^{-1}$)} &
\colhead{($\degr$)} &
\colhead{(km~s$^{-1}$)} &
\colhead{(km~s$^{-1}$)} &
\colhead{(mJy~kpc$^{-2}$)} &
\colhead{(kpc)}}
\startdata
Disk & 02:26:01.87550(12) & $+$03:02:59.2448(20) & 6.54053(3) & $199 \pm 2$ & --- & $12.1_{-  2.9}^{+  3.7}$ & $320_{-70}^{+100}$ & $68.5 \pm 1.0$ & $6.16 \pm 0.12$ & 0.822(14)\\
Warped Disk & 02:26:01.87547(10) & $+$03:02:59.2466(20) & 6.54052(3) & $226 \pm 3$ & $-24.8_{-2.1}^{+1.9}$ & $40.4 \pm 1.3$ & $116 \pm 3$ & $66.4 \pm 1.0$ & $6.56 \pm 0.13$ & 0.888(15)
\enddata
\tablecomments{P.A.$_{1}$ denotes the linear slope in the position angle of the major axis for the warp (see Section \ref{sec:kinematics}). The remainder of the parameters are described in detail in \citep{Neeleman2021}. We note that $I_0$ is the flux density of the central pixel, and is therefore dependent on the pixel scale of the data cube.}
\end{deluxetable*}

\end{document}